\documentclass[10pt, a4paper, twocolumn]{article}

\usepackage[english]{babel} % English language hyphenation

\usepackage{microtype} % Better typography

\usepackage{amsmath,amsfonts,amsthm} % Math packages for equations

\usepackage[svgnames,table]{xcolor} % Enabling colors by their 'svgnames'

\usepackage[hang, small, labelfont=bf, up, textfont=it]{caption} % Custom captions under/above tables and figures

\usepackage{booktabs} % Horizontal rules in tables

\usepackage{lastpage} % Used to determine the number of pages in the document (for "Page X of Total")

\usepackage{graphicx} % Required for adding images

\usepackage{enumitem} % Required for customising lists
\setlist{noitemsep} % Remove spacing between bullet/numbered list elements

\usepackage{sectsty} % Enables custom section titles
\allsectionsfont{\usefont{OT1}{phv}{b}{n}} % Change the font of all section commands (Helvetica)

\usepackage{geometry} % Required for adjusting page dimensions

\usepackage[T1]{fontenc} % Output font encoding for international characters
\usepackage[utf8]{inputenc} % Required for inputting international characters

\usepackage{XCharter} % Use the XCharter font

\usepackage{fancyhdr} % Needed to define custom headers/footers
\fancypagestyle{firstpage}{ % Page style for the first page with the title
	\fancyhf{}
}

\newcommand{\authorstyle}[1]{{\usefont{OT1}{phv}{b}{n}\color{Black}#1}} % Authors style (Helvetica)

\newcommand{\institution}[1]{{\footnotesize\usefont{OT1}{phv}{m}{sl}\color{Black}#1}} % Institutions style (Helvetica)

\usepackage{titling} % Allows custom title configuration

\newcommand{\HorRule}{\color{DarkGoldenrod}\rule{\linewidth}{1pt}} % Defines the gold horizontal rule around the title

\pretitle{
	\vspace{-50pt} % Move the entire title section up
	\HorRule\vspace{10pt} % Horizontal rule before the title
	\fontsize{20}{28}\usefont{OT1}{phv}{b}{n}\selectfont % Helvetica
	\color{DarkRed} % Text colour for the title and author(s)
}

\posttitle{\par\vskip 15pt} % Whitespace under the title

\newcommand{\subtitle}[1]{%
  \posttitle{%
  	\par\vskip0.5em
    \Large#1
    \vskip0.5em}%
}

\preauthor{} % Anything that will appear before \author is printed

\postauthor{ % Anything that will appear after \author is printed
	\vspace{10pt} % Space before the rule
	\par\HorRule % Horizontal rule after the title
	\vspace{8pt} % Space after the title section

}

\usepackage{lettrine} % Package to accentuate the first letter of the text (lettrine)
\usepackage{fix-cm}	% Fixes the height of the lettrine

\usepackage{xstring} % Required for string manipulation

\usepackage[english]{babel}
\usepackage{amsmath}
\usepackage{bbold}
\usepackage{graphicx}
\usepackage{mdframed}
\usepackage{amssymb}
\usepackage{cite}

\usepackage{tikz}

\newcounter{movie}
\newenvironment{movie}[1][]{\refstepcounter{movie}\par\medskip
   \textbf{Movie~\themovie. #1} \rmfamily}{\medskip}

\title{The power of the AC-DC circuit}
\subtitle{Operating principles of a simple multi-functional \\transcriptional network motif}
\author{\authorstyle{Ruben Perez-Carrasco\textsuperscript{1}, Chris P. Barnes\textsuperscript{2,3}, Yolanda Schaerli\textsuperscript{4},\\ Mark Isalan\textsuperscript{5}, James Briscoe\textsuperscript{6}, Karen M. Page\textsuperscript{1}}\\
\institution{\textsuperscript{1} Department of Mathematics, University College London, Gower Street, WC1E 6BT, London, UK\\
\textsuperscript{2} Department of Cell and Developmental Biology, University College London, Gower Street, WC1E 6BT, London, UK\\
\textsuperscript{3} Department of Genetics, Evolution and Environment, University College London, Gower Street, WC1E 6BT, London, UK\\
\textsuperscript{4} Department of Fundamental Microbiology, University of Lausanne, Biophore Building, 1015 Lausanne, Switzerland\\
\textsuperscript{5} Department of Life Sciences, Imperial College London, SW7 2AZ, London, UK\\
\textsuperscript{6} The Francis Crick Institute, 1 Midland Road, NW1 1AT London, UK
}}

\begin{document}
\maketitle
\date{\today}
\section*{Abstract}
\begin{mdframed}[backgroundcolor=black!20,linewidth=0,leftmargin=-0.5cm,rightmargin=-0.3cm]
\textbf{Genetically encoded regulatory circuits control biological function. A major focus of systems biology is to understand these circuits by establishing the relationship between specific structures and functions.  Of special interest are multifunctional circuits that are capable of performing distinct behaviors without changing their topology. A particularly simple example of such a system is the AC-DC circuit. Found in multiple regulatory processes, this circuit consists of three genes connected in a combination of a toggle switch and a repressilator. Using dynamical system theory we analyze the available dynamical regimes to show that the AC-DC can exhibit both oscillations and bistability. We found that both dynamical regimes can coexist robustly in the same region of parameter space, generating novel emergent behaviors not available to the individual subnetwork components. We demonstrate that the AC-DC circuit provides a mechanism to rapidly switch between oscillations and steady expression and, in the presence of noise, the multi-functionality of the circuit offers the possibility to control the coherence of oscillations. Additionally, we provide evidence that the availability of a bistable oscillatory regime allows the AC-DC circuit to behave as an excitable system capable of stochastic pulses and spatial signal propagation. Taken together these results reveal how a system as simple as the AC-DC circuit can produce multiple complex dynamical behaviors in the same parameter region and is well suited for the construction of multifunctional synthetic genetic circuits. Likewise, the analysis reveals the potential of this circuit to facilitate the evolution of distinct patterning mechanisms.}
\end{mdframed}

% \affil[a]{Department of Mathematics, University College London, Gower Street, London WC1E 6BT}
% \affil[b]{Department of Cell and Developmental Biology, University College London, Gower Street, WC1E 6BT, London, UK}
% \affil[c]{Department of Genetics, Evolution and Environment, University College London, Gower Street, WC1E 6BT, London, UK}
% \affil[d]{Department of Fundamental Microbiology, University of Lausanne, Biophore Building, 1015 Lausanne, Switzerland}
% \affil[e]{Department of Life Sciences, Imperial College London, London SW7 2AZ, UK}
% \affil[f]{The Francis Crick Institute, 1 Midland Road, NW1 1AT London, UK}

\thispagestyle{firstpage} % Apply the page style for the first page (no headers and footers)
\section*{Introduction}

Genetic circuits regulate biological functions in a variety of contexts, ranging from embryonic development to tissue homeostasis. Accordingly, the analysis of the repertoire of functions performed by genetic circuits is central to systems biology. In some cases there is a direct relationship between the structure and operation of a circuit, such that the function -- the dynamical behavior -- of a circuit is evident from its topology. This has led to the classification of motifs or subnetworks based on topology and motivated the design and fabrication of artificial circuits with functions that include toggle switches, band-pass filters, memory devices, logic gates, and oscillators \cite{Gardner2000,Elowitz2000,Basu2005,Sohka2009,Ajo-franklin2007,Siuti2013}. Engineered versions of these circuits have been deployed to perform computation, screen for drugs, detect and treat diseases \cite{Daniel2013,Rubens2016,Xie2016}. 

Nevertheless, there is not always a one-to-one correspondence between topology and behavior. This is apparent from the analysis of even small circuits. In these cases, a small modification to such a circuit, for example the change in strength of interactions between components, leads to a qualitative change in the behavior of the circuit \cite{Jia2017,DelVecchio:2008gy, Jayanthi:2013do,Tan:2009de,Prindle:2014hj,Ingram2006}. Far from being nuisance, this has led to the concept of multi-functionality \cite{Jimenez2017, Purcell2011} -- circuits capable of qualitatively different outputs in a reduced parameter range. This poses the challenge of defining and predicting circuit behavior and emphasizes the importance of understanding the mapping between the topology of a genetic network and its dynamical behavior.

Understanding the minimal parameter changes necessary to elicit alternative behaviors from a multi-functional circuit provides insight into changes in behavior during gene network evolution and could be exploited for the engineering of circuits for synthetic biology tasks. Several studies have shed light on this problem through extensive numerical exploration of small networks targeting a specific function \cite{Cotterell2010,Woods2016,Jimenez2017,Otero-Muras2016}. The insight of such studies is often obtained after the analysis and classification of the successful topologies in terms of the landscape of the corresponding dynamical system -- sometimes called, the geometrical landscape. The use of this dynamical landscape is key to revealing how different behaviors emerge, contributing to a better understanding of the mapping of topology to function \cite{Jia2017,Strelkowa2010,Suel2006,Jaeger2014,Verd2014} .

Distilling minimal easy-to-engineer networks capable of specific functions is of paramount importance for engineering circuits for synthetic biology tasks  \cite{Schaerli2014,Purcell2010,Chau2012}. Cellular resources are scarce and implementing complex behaviors within cells requires an efficient and judicious use of these \cite{Carrera:2011fd, Mather:2013hn, Cookson:2011il}. Metabolic load affects gene expression through growth dependent effects \cite{Klumpp:2009vu, Scott:2010cx,Cardinale:2013bt} and its reduction has become a major design objective \cite{Ceroni:2015ep,Borkowski:2016jz}. Minimal multi-functional circuits might offer a potential route to this goal. It is therefore critical to find and understand the behaviors and emergent properties that can be encoded in a reduced gene circuitry. Theoretical and computational analyses have revealed that merely combining modules with different functions does not necessarily lead to additive outcomes. Conversely, in many cases, topologies capable of multi-functional behavior can not be explained simply as the overlap of two or more submodules \cite{Jimenez2017}.  A deeper understanding of the dynamics of multi-functional circuits is needed. As there is ample evidence that real biological systems exploit multi-functionality \cite{Verd2016}, designing and investigating such circuits is likely to shed light on biological processes \cite{Mathur2017}.  
 
\begin{figure}[th!]
\centering
\includegraphics[width=1.0\columnwidth]{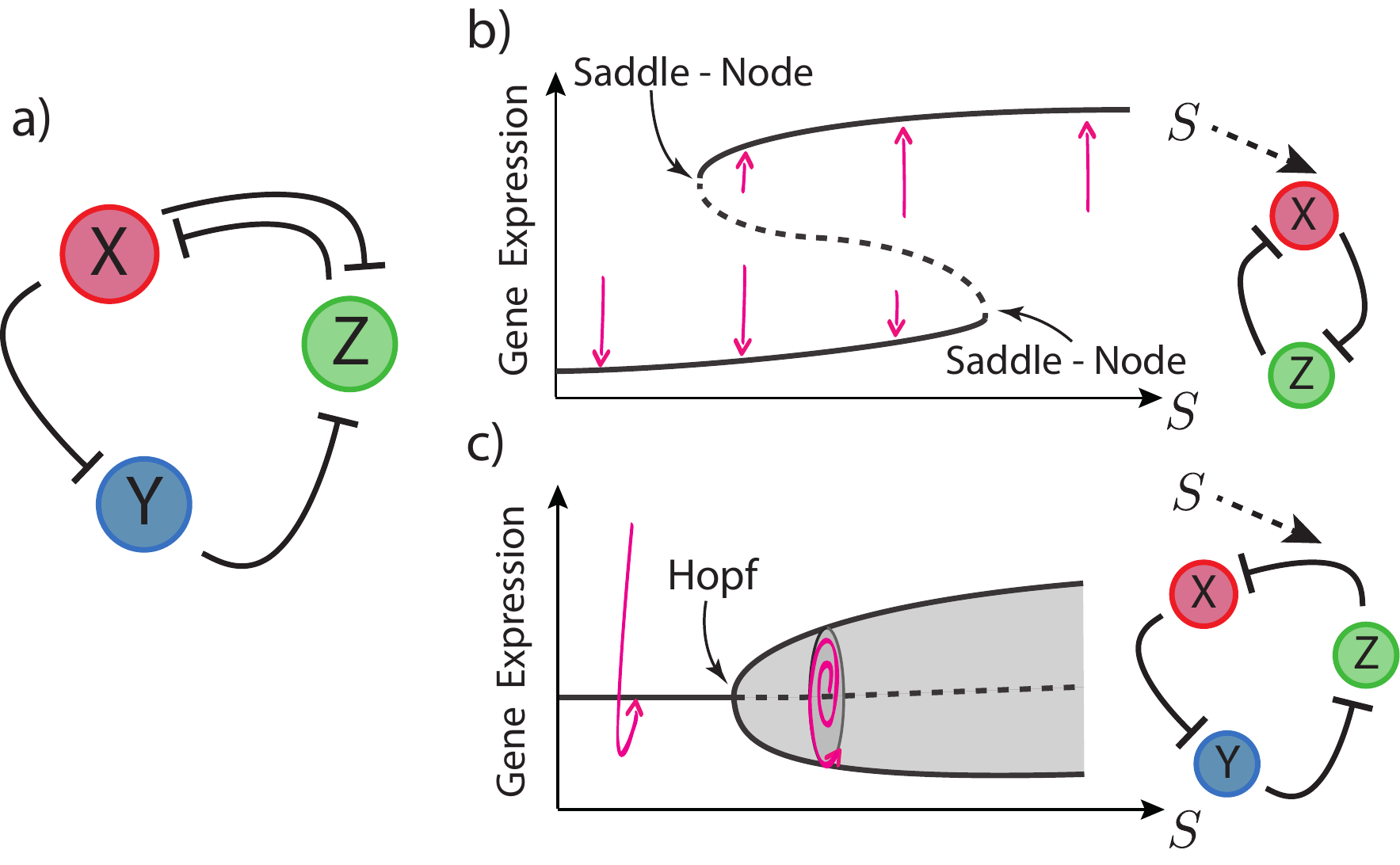}
\caption{\label{fig.schemeAC-DC} \textbf{The AC-DC is the combination of a bistable switch and a repressilator} a) AC-DC regulatory circuit. b) Bifurcation diagram and network diagram for the bistable switch controlled by a signal $S$. The two saddle-node bifurcations position the stability range for the two stable solutions (\emph{black solid lines}). There is bistability for intermediate signals, where both stable solutions are separated by an unstable steady state (\emph{dashed line}). Transient trajectories (\emph{pink arrows}) are sketched showing the dynamical effect of the steady states.  c) Bifurcation diagram and network diagram for the repressilator under a change of parameters controlled with an external signal $S$. The change in behavior is a Hopf bifurcation where a stable spiral (damped oscillations) (\emph{black solid line}) becomes unstable (\emph{dashed line}) giving rise to stable oscillations of growing amplitude. The two possible oscillatory transients are sketched (\emph{pink arrows}).} 
\end{figure}

One attractive candidate for studying the coexistence and emergence of behavior in a multi-functional minimal network is the AC-DC circuit \cite{Panovska-Griffiths2013,Balaskas2012}. Composed of two well known subnetworks, the repressilator \cite{Elowitz2000,Purcell2010} and the toggle switch \cite{Sokolowski2012,Gardner2000} (Fig. \ref{fig.schemeAC-DC} a)) the AC-DC circuit takes its name by analogy to Alternate (AC) and Direct (DC) Current, since it is capable of generating oscillatory (AC) and multistable (DC) behavior \cite{Panovska-Griffiths2013}. The AC-DC circuit was originally observed in the patterning of progenitors in the vertebrate neural tube \cite{Balaskas2012}, where it is proposed to exhibit  the DC behavior. Later, theoretical analysis revealed the potential for this circuit to generate oscillations \cite{Panovska-Griffiths2013}.  Strikingly, Jaeger and colleagues have proposed that the gap gene system, which patterns the anterior-posterior axis of the \emph{Drosophila melanogaster} embryo, is composed of three linked AC-DC circuits, two of which operate in the DC regime and one in the oscillatory, AC, mode \cite{Verd2016}.  

The two subcomponents of the AC-DC circuit, the toggle switch and the repressilator, have been intensively studied, separately. Toggle switches consist of the cross-repression between the determinants of different cellular states and result in a definite choice between two outcomes. When controlled by an external signal, the toggle switch is able to produce a sharp transition between the steady states at a precise signal level (Fig. \ref{fig.schemeAC-DC} b)) \cite{Sokolowski2012,Gardner2000}. From a dynamical systems point of view, the sharp switch-like transition is the result of two saddle-node bifurcations. Each of these is characterized by the abrupt appearance of a stable and unstable steady expression state for a small change in the input signal. A consequence of this dynamical scenario is that for a range of values of signal, both states are available and the expression state is determined by the initial gene expression. Additionally, in the presence of noise, there is the possibility of switching between the stable states \cite{Song2010,Tian2006,Perez-Carrasco2016,Frigola2012}. 

The second component of the AC-DC circuit, the repressilator, comprises the sequential repression of three genes. In contrast to the toggle switch this provides a negative feedback loop that promotes stable oscillations. The amplitude and period of these depend on the parameters of the system \cite{Elowitz2000,Purcell2010}. Changes in these parameters can lead to the disappearance of the oscillations through a Hopf bifurcation, in which the orbit in the expression space (limit cycle) shrinks, giving rise to a steady expression state. Hence coupling key parameters to an external signal can result in the repressilator becoming a switchable genetic oscillator, a property that has been extensively computationally explored \cite{Buzzi2015,BUSE2009,Purcell2010}.

In this manuscript we characterize the functions of the AC-DC circuit by analyzing the phase portrait of the dynamical system. We find that oscillations and stable expression can coexist in a large region of the parameter space and we explore the implications of this coexistence. This reveals emergent behaviors not available to the repressilator or the toggle switch individually, that allow the circuit to be used to establish coherent or incoherent oscillations. Additionally, we demonstrate that, with the addition of noise, the AC-DC circuit functions as an excitable system capable of stochastic pulses and spatial signal propagation. 

\section*{The model}

The expression dynamics of the AC-DC circuit can be described taking into account the production and degradation of each gene ($X$,$Y$,$Z$), where transcription processes are assumed to be faster than translation \cite{Panovska-Griffiths2013}. The production rate of each gene is regulated by the genetic interactions in the network and the inductive signal ($S$) that controls the behavior of the network and activates genes $X$ and $Y$,
\begin{eqnarray}
\dot X&=&\frac{\alpha_X+\beta_X S}{1+S+(Z/z_X)^{n_{zx}}}- X, \nonumber\\
\dot Y&=&\frac{\alpha_Y+\beta_Y S}{1+S+(X/x_Y)^{n_{xy}}}-\delta_Y Y, \label{eq:hillfunction}\\
\dot Z&=&\frac{1}{1+(X/x_Z)^{n_{xz}}+(Y/y_Z)^{n_{yz}}}-\delta_Z Z. \nonumber
\end{eqnarray}

Here all the variables and parameters are non-dimensional (see SI). The non-dimensional basal production rates $\alpha_X$ and $\alpha_Y$ are relative to the basal production rate of gene Z. The signal induction is controlled by parameters $\beta_X$ and $\beta_Y$, while the strength and shape of the repressions are controlled by the non-dimensional factors $z_X$, $x_Y$, $x_Z$, and $y_Z$ and the exponents $n_{zx}$, $n_{xy}$, $n_{xz}$, and $n_{yz}$. Finally, the rates $\delta_Z$ and $\delta_Y$ are the relative degradation rates to the degradation of gene $X$. Similarly, the time is measured in units of time of the degradation rate of protein $X$.

The use of non-dimensional parameters allows for the study of the minimal independent set of parameters required to define the possible different dynamics of the circuit, maximizing the information obtained for any parameter fitting of the model. In the present case we are interested in finding a global behavior of the AC-DC circuit without overfitting. For this reason we performed a minimal parameter exploration only looking for behaviors that exhibit a transition from non-oscillatory to oscillatory behavior through a change in the signal. Additionally, we required the Hill exponents to be as low as possible to avoid numerical artifacts due to high nonlinearities. This would also ensure a set of parameters achievable in synthetic circuits. 

The parameter exploration was performed using Approximate Bayesian Computation (ABC) \cite{Liepe2010,Liepe2014} and gave as a result the distribution of parameters necessary for observing a tunable oscillator in the AC-DC circuit.  The resulting parameters (Table \ref{tab:parameters}) return a consistent relationship between the parameters for different target optimizations (see SI). Namely, the basal production rate of the different genes has a marked hierarchy with $Z$ being the largest, followed by gene $X$ and $Y$. By contrast, both signal activation strengths are similar ($\beta_X\simeq\beta_Y$).  Additionally, the strongest repression is that of X from Z, while the weakest is its reciprocal, from X to Z. The repressions unique to the repressilator, $x_y$ and $y_z$ are in between these values. All the optimizations returned a difference of at least one order of magnitude between the different repression magnitudes ($z_x<x_y<y_z<x_z$), with clear correlations between them. Notably, a similar degradation rate was observed for all three proteins $\delta_Y\simeq\delta_Z\simeq 1$. Finally, no oscillations were found when Hill exponents $n=2$ were used, but a small increase of only one of the Hill exponents provided sufficient non-linearity to observe oscillations.

In addition to the deterministic model, it is also informative to test the behavior of the AC-DC circuit subjected to molecular intrinsic noise derived from the deterministic equations [\ref{eq:hillfunction}] as Chemical Langevin Equations \cite{Gillespie2000} (see SI). The inclusion of intrinsic noise has two purposes, it shows the robustness of some of the functionalities to fluctuations, while revealing new phenomena not available in a deterministic scenario.
  
\section*{AC-DC circuit shows bistability between oscillations and steady expression}

\begin{figure}[t]
\centering
\includegraphics[width = 0.9\columnwidth]{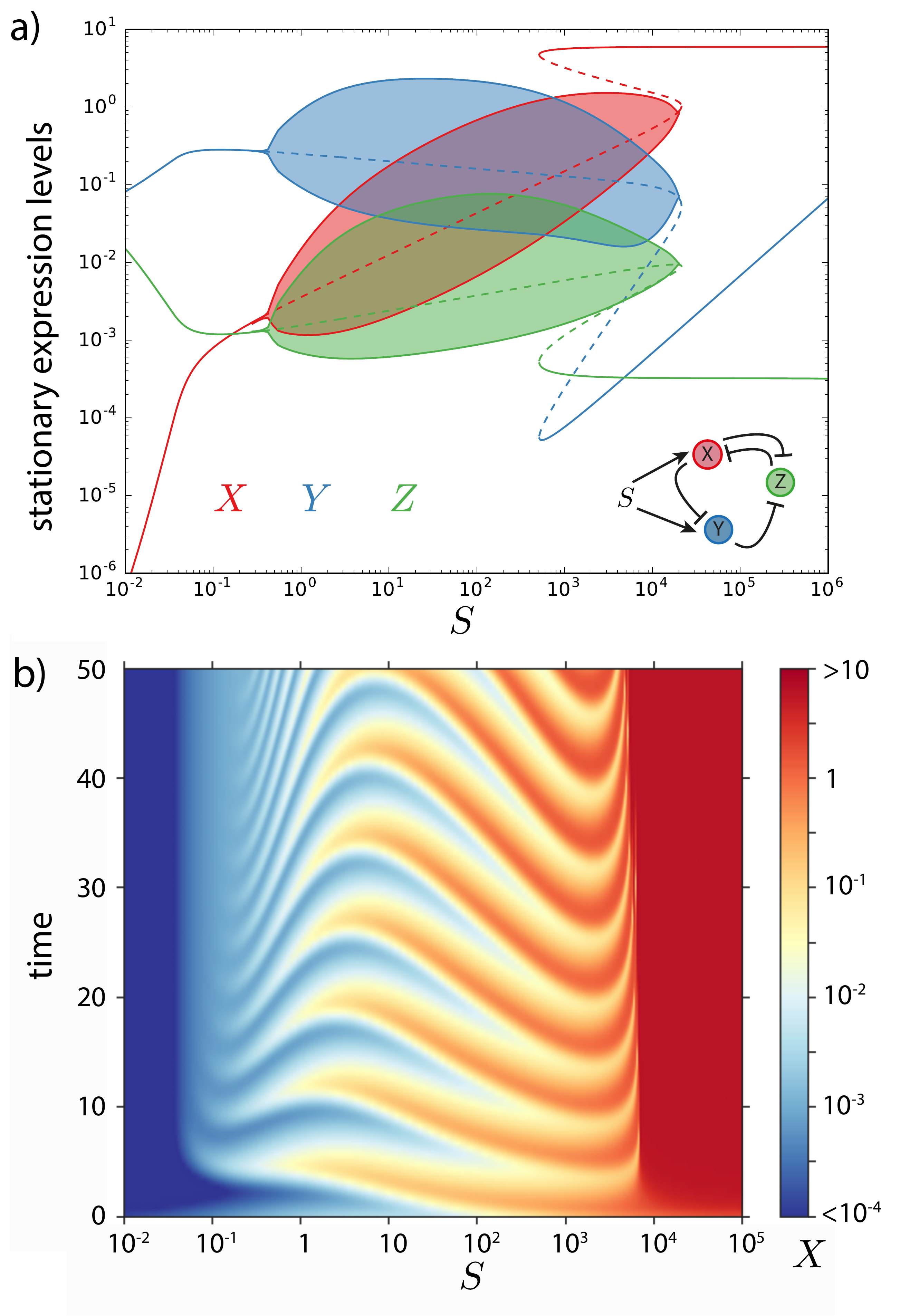}
\caption{\label{fig:stability} \textbf{Dynamical regimes of the AC-DC} a) Stability diagram showing the available steady states for each value of the signal $S$ for the parameters of Table \ref{tab:parameters}. Solid lines show stable steady states, dashed lines show unstable states. Oscillatory states have an amplitude spanning the shaded areas. Bifurcation diagram was obtained using integration and continuation techniques \cite{Clewley2012}. b) Expression of gene $X$ in time for different Signal levels exhibiting three different dynamical regions. Initial condition is $X = 0, Y = 0, Z = 0$.}
\end{figure}

Analysis of the bifurcation diagram of the circuit reveals a mixture of the bistability from the toggle switch and the oscillatory behavior of the repressilator. The Hopf bifurcation by which the oscillations arise in the repressilator transforms one of the stable states of the bistable switch into an oscillatory state that can coexist for a certain signal range with the other stable state (Fig. \ref{fig:stability} and movies \ref{movie:phasecolor}, \ref{movie:phasechannels}, \ref{movie:phasetimegradient}). Hence, for a given value of signal $S$ both behaviors (oscillatory or stable expression) are possible. The chosen state will depend on the history of the system $S$, \emph{i.e.} the system displays hysteresis. Surprisingly this behavior was present in 80\% of the optimized parameter sets even though the parameter search optimization did not score for any kind of bistability. This suggests that it is a robust behavior arising from the network topology. 

Overall, examining the bifurcation diagram shows that the behavior consisted of four different dynamical regimes for different signal ranges. For low values of the signal, there is only one possible steady state with low expression of gene X, resulting from a low activation of the promoters X and Y by the signal. As the signal increases, the system starts to oscillate through a Hopf bifurcation, with oscillations of small amplitude that increases with the amount of signal. For larger values of the signal a new stable state with high expression of $X$ becomes available through a saddle-node bifurcation. This new state appears away from the limit cycle (steady oscillatory trajectory) without affecting it, giving rise to the bistable regime between oscillations and constant expression.

\begin{figure*}[!h]
\centering
\includegraphics[width = 1.0\textwidth]{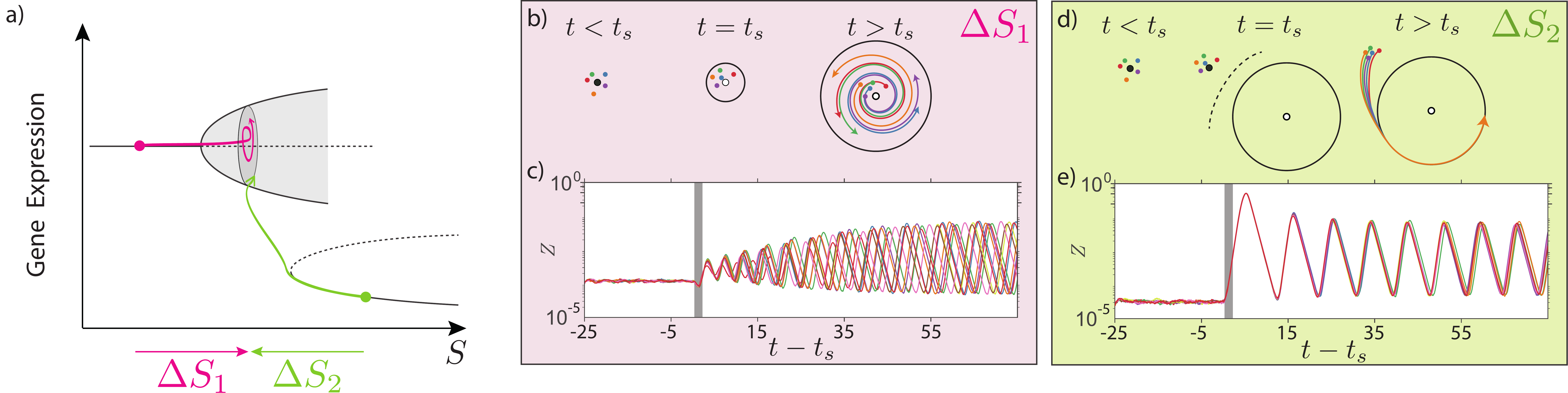}
\caption{\label{fig:coherence}\textbf{AC-DC allows the control of oscillation coherence between different cells.} a) Schematic showing the two possible transitions towards the limit cycle. b) Oscillations arising through the Hopf bifurcation ($\Delta S_1$) are incoherent. Diagrams show steady states and transients in the genetic expression plane.  Initially, there is only one stable state (\emph{solid black circle}) of constant expression, the genetic expression of different cells (\emph{colored circles}) is determined by this stable state. After the signal is increased at $t=t_s$, the steady state becomes unstable giving place to an unstable spiral center (\emph{solid white circle}) and a stable limit cycle (\emph{black circumference}). The resulting dynamical behavior for the different cells (\emph{coloured arrows}) follow a spiral transient towards the limit cycle. c) Simulations of $\Delta S_1$ show the appearance of oscillations that lose their coherence by increasing the signal from $S=0.1$ to $S=100$ at $t=t_s$ for $\Delta t = 2$ (\emph{grey shaded area}), $\Omega=10^6$. d) Oscillations through the saddle-node bifurcation ($\Delta S_2$) are coherent. Diagrams show steady states and transients in the genetic expression plane.  Initially, expression of cells (\emph{coloured circles}) are found in a stable state (\emph{solid black circle}) of constant expression. After the signal is increased at $t=t_s$, the steady state disappears from the plane  (\emph{solid white circle}) and the only attractor available is the limit cycle (\emph{black circumference}) that imposes a fast genetic expression transient towards the stable oscillations (\emph{coloured lines}). e) Simulations of $\Delta S_2$ show the appearance of coherent oscillations by decreasing the signal from  $S=10^5$ to $S=100$ at $t=t_s$ for $\Delta t = 2$ (grey shaded area),  $\Omega=10^6$.}
\end{figure*}
For large values of the signal the oscillations disappear. The bifurcation analysis indicated that two different mechanisms could be responsible for this. On one hand, a Hopf bifurcation may arise collapsing the limit cycle before the second saddle-node occurs (Fig. \ref{fig:stability}). On the other hand, the oscillatory state can collide with the unstable steady state produced by the saddle-node bifurcation, resulting in a homoclinic bifurcation. This gives rise to a regime in which even though the oscillatory state is not stable, it readily generates oscillatory transients towards the steady state (Fig. \ref{fig:homoclinic}). We will consider the first case for the rest of this study, nevertheless, all the behaviors described herein are independent of the mechanism by which the oscillations disappear.

\section*{Coherent or incoherent oscillations}

One way of inducing oscillations through a change in signal is to increase the signal from low levels to a level above the Hopf bifurcation threshold ($\Delta S_1$). In addition, the coexistence of oscillations with saddle-node bifurcations allows for an alternative way to initiate oscillations. Starting from the stable expression state achieved at high signal levels, the oscillatory state can be reached by reducing the signal below the saddle-node bifurcation ($\Delta S_2$) (see Fig. \ref{fig:coherence}). Whereas in the first case a small limit cycle arises around the initial steady state, in the second case, a large amplitude limit cycle is already present within the dynamical landscape when the bifurcation takes place. These differences result in different dynamical transients towards the oscillatory state.

In order to test these differences we performed simulations of the stochastic model starting at low or high signal and ending at the same intermediate signal value. Results show that in the first scenario -- increasing signal from a low value -- gives rise to asynchronous oscillations in a population of cells. Small differences in the initial phase are amplified over time. By contrast, the second scenario -- decreasing signal from a high level -- induces coherent oscillations in the face of noise (Fig. \ref{fig:coherence} and Movies \ref{movie:coherenceHopf} and \ref{movie:coherenceSN}).

This difference in behavior is a consequence of the different initial gene expression states in relation to oscillatory spiral center. In a Hopf bifurcation, oscillations arise through an attracting spiral losing its stability and becoming a repulsing spiral. Hence, oscillations originating from a Hopf bifurcation start their transient close to the unstable spiral centre and a small variation in the initial condition can lead to a substantial difference in the final oscillation phase. Small initial differences are amplified resulting in lack of coherence of oscillations for a population of cells undergoing the bifurcation. On the contrary, cells passing through the saddle-node bifurcation towards the limit cycle do so at expression levels that are far from those associated with the attracting oscillatory regime. Consequently, they have the same initial phase and stochastic trajectories are canalized together towards the oscillatory state. In this way the AC-DC circuit, for a single set of parameters, offers the possibility to establish either coherent or incoherent oscillations in a population by choosing the appropriate signal transient. Importantly, the second mechanism is not available in the original repressilator since it requires the bistability provided by the toggle switch.
 
In addition to the synchrony of response, it is important to note that the saddle-node bifurcation also allows the rapid establishment of constant amplitude oscillations. Thus, the AC-DC system offers a fast mechanism to turn on and off the oscillations that is not a feature of the repressilator. Previous studies propose robust switching exploiting quasi-stable oscillatory transients in repressilators with an even number of repressions \cite{Strelkowa2010}. In contrast the AC-DC circuit exhibits the benefits of both systems, the robustness of oscillating with a stable limit cycle, and the fast switchability, in this case provided by a bifurcation occurring far from the central unstable spiral.

\begin{figure}[]
\includegraphics[width = 0.9\columnwidth]{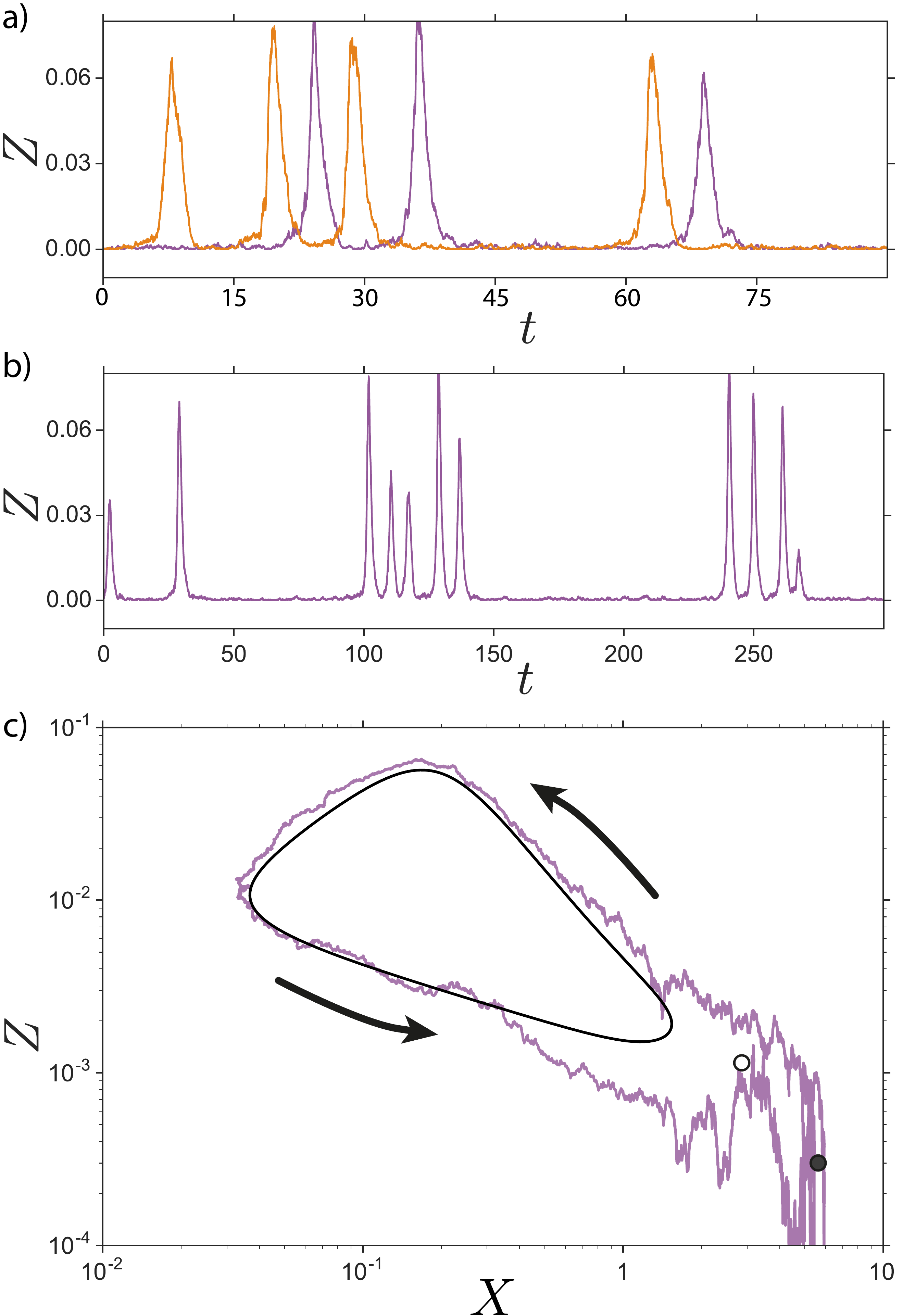}
\caption{\label{fig:excitation} \textbf{AC-DC circuit behaves as a type II excitable system.} a) Pulsing time traces of genetic expression for two different realizations (different colors). $S=1000$, $\Omega=1000$ b) Train of spikes of genetic expression $S=1000$, $\Omega=3000$ c) Expression levels of genes X and Z during a pulse. Stochastic fluctuations drive the system far from the steady state (\emph{black circle}) past the unstable steady state (\emph{white circle}) driving the system around the limit cycle (\emph{black orbit}). }
\end{figure}

\begin{figure}
\includegraphics[width = 0.9\columnwidth]{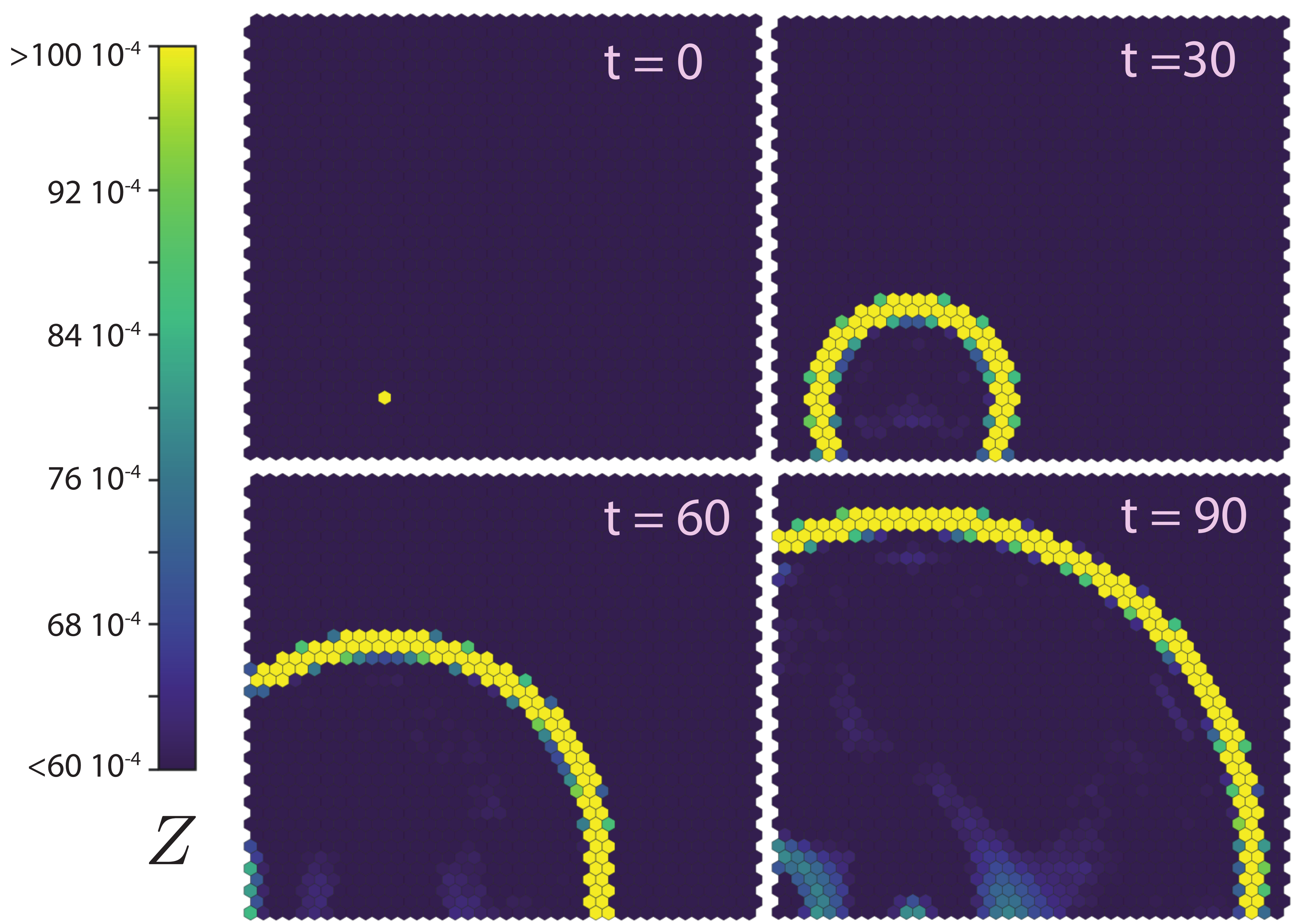}
\caption{\label{fig:propagation} \textbf{AC-DC pulse can be used as a spatial signal propagation} Array of cells containing the AC-DC circuit. Proteins X, Y and Z can diffuse intercellularly. Cells are under a signal $S = 1000$ (bistable regime) are initially at the constant expression steady state except one cell that is perturbed away from the steady state. $\Omega = 10^{5}$, $D = 0.1$ (see more details in SI).}
\end{figure}

\section*{The AC-DC circuit shows excitability}

The long-term behavior of the deterministic system in the bistable zone of the AC-DC circuit is determined by the initial conditions of the system. The set of initial conditions that are attracted to each of the two possible stable states, are their respective basins of attraction. Intrinsic fluctuations in the expression levels allow the system to explore the basin of attraction or even to cross between basins. This has been related with noise-induced transitions between different cellular states \cite{Song2010,Jia2017}.  In the case of the AC-DC circuit, a switching of behavior between oscillations and constant expression can result from intrinsic noise without the necessity to change the external signal. The frequency of such switching will depend on the geometry of the basin together with the level of intrinsic noise. 

Switching between states is not equally probable for all levels of gene expression \cite{DeLaCruz2017}. In particular, once the system jumps towards an oscillatory state, at least one excursion around the limit cycle is required before it can return to the constant expression state (Fig. \ref{fig:excitation}). Such an excursion results in the amplification of a transient fluctuation. Additionally, this excursion entails a refractory period during which the system cannot be triggered again, until the full cycle is finished. This pulsatile behavior reveals the excitable nature of the AC-DC circuit. Excitability has been found in other genetic systems where expression pulses are beneficial for the biology of the cell \cite{Suel2006,Levine2013}, usually resulting from incoherent feedbacks. Incoherent feedback in the AC-DC circuit arises from the superposition of its two subcircuits, the bistable switch (positive feedback) and the repressilator (negative feedback). The frequency of the resulting pulses will depend on the noise intensity and value of the signal. Furthermore, for signals where the oscillatory state is more stable, the excitation pulses could lead to trains of pulses, arising from the trapping of the system in the oscillatory state for more than one period. This is similar to the spike trains observed in neuronal activity \cite{Lindner2004}. We observed these spike trains by changing the levels of intrinsic noise (Fig. \ref{fig:excitation}) through the noise intensity $\Omega$ (see SI)\cite{Gillespie2000,Perez-Carrasco2016}. 

The excitable nature of the AC-DC circuit might also be relevant to other functions, such as signal propagation in a tissue. A single cell signalling to neighboring cells can be excited to undergo a pulse that will in turn excite neighboring cells and so on. The intensity, transient and refractory period of the pulse not only contribute to the excitation of neighboring cells but also inhibit the reactivation of the recently excited cells, resulting in a spatially propagating pulse over the tissue. In order to test this possibility, we performed a series of numerical assays in a simulated tissue where one or more of the proteins forming the AC-DC circuit diffuse between cells. To initiate the system, bistable cells are set in the constant stable expression state. In this scenario, the induced excitation of one cell leads to a propagating front in which cells are excited sequentially returning afterwards to the initial constant expression state (Fig. \ref{fig:propagation} and Movie \ref{movie:propagationnice}). During this period of time, the transient expression along the limit cycle is high enough to deliver the pulse to the neighboring cells. The width and velocity of the propagating front can be controlled with the noise intensity or the diffusion coefficient, as well as the signaling mechanism between cells. Additionally, for a high enough level of noise, spontaneous propagation waves can also occur, as well as dynamical patterning of the system (see Movies \ref{movie:propagationnice}, \ref{movie:propagationquick}, \ref{movie:propagationdirty}, and \ref{movie:propagationsingle} ). Similar results would be expected with more elaborate signalling pathways in which transmembrane receptors are involved in the transmission of the signal \cite{Mathur2017,Jimenez2017,Formosa-Jordan2012}.   

\section*{Discussion}

We have explored the behavior of the multifunctional AC-DC circuit, showing that the coexistence between bistability and oscillations elicit novel dynamics that are unavailable to either of its constituent parts. These provide a mechanism to rapidly switch between a rhythmic and steady expression regime and suggests a way in which coherent oscillations can be induced into an ensemble of otherwise noisy oscillators. Additionally, we demonstrate that the circuit can behave as an excitable system, capable of stochastic pulsing. Importantly, these behaviors are accessible for the same range of parameters, supporting the idea that the circuit represents a versatile genetically encoded network well suited for a range of functions.

Although not apparent from the structure of the network, the various functions of the AC-DC circuit become evident from an inspection of the geometry of its dynamical landscape. As has been shown in previous studies, the position and nature of the attractors of the system provide substantially greater insight than a simple analysis of the topology of the network\cite{Jimenez2017,Cotterell2010,Suel2006,Strelkowa2010,Jia2017,Jaeger2014,Verd2014}. For the AC-DC circuit, the shape of the dynamical landscape is created by the combination of saddle-node bifurcation arising from the toggle switch and the Hopf bifurcation of the repressilator. Moreover, new transitions, such as the sudden destabilization of oscillations through the homoclinic bifurcation, can be anticipated from examining the structure of the dynamical landscape. This highlights the application and importance of dynamical systems tools for identifying and explaining the behavior of even relatively simple circuits. It also raises the possibility that similiar behaviors, including bistability between a limit cycle and a steady state, might be present in other circuits composed of incoherent feedbacks. 

Of particular interest for the AC-DC dynamics are the regions of multistability and the ability to control switching between the different behaviors by changing a single external signal. The availability of wide regions of parameter space in which these behaviors take place makes the AC-DC circuit an attractive target for synthetic biology. For example, generating synchronized ensembles of oscillators has been a challenge. Considerable efforts have been made either by engineering away noise or relying on quorum sensing\cite{Kobayashi:2004cv,Kuznetsov:2006gx,Danino2010,Nikolaev:2016cv,Potvin-Trottier2016}. The bistable oscillation of the AC-DC circuit offers an alternative strategy for achieving coherent oscillations in response to a triggering signal and this might be extendable to other network topologies displaying bistability. Alternatively, for tasks in which incoherent oscillations are desired, the AC-DC circuit retains the potential to initiate oscillations via a conventional Hopf bifurcation. 

Pulsatile excitations are also a property exploited in several biological situations. The behavior is reminiscent of bet-hedging strategies that have been proposed to optimize responses to external inputs or maximize the use of limited resources by controlling the time at which nutrient demanding physiological processes occur \cite{Levine2013,Suel2006,Liu2017}. From this perspective, the AC-DC circuit provides a mechanism to explore different excitable regimes just by changing the external signal without the need to control the levels of intrinsic noise by changing the copy number or degradation rates of the system \cite{Hilborn2012,Niederholtmeyer2015}. Moreover, the AC-DC circuit exhibits pulsatile properties not found previously. In other circuits the excitability arises from unstable excitable transients or through a subcritical Hopf bifurcation \cite{Suel2006,Hilborn2012}. In the AC-DC circuit, the separation of the saddle-node bifurcation that initiates oscillations from the amplitude of the limit cycle allows for parameterizations in which both properties of the bistable region can be tuned independently to control the different features of the pulses. Similar distinctions are found in excitable systems such as those associated with neuronal action potentials \cite{Izhikevich2000,Lindner2004}, raising the possibility of combining current advances in neuronal networks and excitable media with synthetic genetic circuits.

The AC-DC circuit also offers insight into genetic circuits involved in tissue development. During vertebrate embryogenesis, coordinated gene expression oscillations -- the segmentation clock -- in the posterior cells of the body generate a rhythmic spatial pattern that subdivides the embryonic trunk into morphological segments \cite{Hubaud2014}. This involves a still poorly defined genetic oscillator within posterior cells. Strikingly, individual cells appear to behave as autonomous oscillators that, when isolated from the embryo, produce transient stochastic periods of oscillations \cite{Webb2016}. This behavior features the same properties found in the excitable regime of the AC-DC circuit in  which a limit cycle and steady expression state coexist, suggesting that the AC-DC circuit could provide a model for the process. In a different tissue, the \emph{Drosophila} blastoderm, the dynamics of three linked AC-DC gene circuits have been proposed to characterise the regulatory network that patterns the anterior-posterior axis \cite{Verd2016}. In this case, the dynamical transients of the AC-DC circuit are suggested to tune the position of the boundaries in time. Moreover, the presence of AC-DC dynamics in this gene network has been suggested later on to reconcile differences between short and long germ-band insects \cite{Clark2017}. In short germ band insects, rhythmic expression of genes is associated with the gradual extension of the body axis. By contrast in long germ band insects, such as \emph{Drosophila}, the trunk is patterned simultaneously without cyclic expression of the patterning genes.  The bistability between a stable steady state and a limit cycle and the possibilities to transition smoothly between regimes with a change of one parameter suggests a route for the evolutionary transition of the underlying gene regulatory network. In xhis view, the multifunctionality of the AC-DC circuit contributes to the evolvability of the circuit and exemplifies how the competing demands of biological mechanisms to be both robust and adaptable can be satisfied. Accordingly, studies of genetically encoded circuits such as the AC-DC network provide insight into the design principles of regulatory mechanisms that characterise biology. 

\subsection*{Acknowledgments}

RPC and KMP would like to acknowledge support from the Wellcome Trust (grant reference WT098325MA ). CPB is funded through a Wellcome Trust Research Career Development Fellowship (097319/Z/11/Z). MI is funded by a Wellcome Trust UK New Investigator Award (WT102944). YS acknowledges support by the Swiss National Science Foundation (Ambizione program PZ00P3-148235). JB is supported through the Francis Crick Institute by funding from Cancer Research UK (FC001051), UK Medical Research Council (FC001051), Wellcome Trust (FC001051 and WT098326MA).

\bibliographystyle{pnas-new}
\bibliography{ACDC,ACDCChris} 

\begin{thebibliography}{10}

\bibitem{Gardner2000}
Gardner TS, Cantor CR, Collins JJ (2000) {Construction of a genetic toggle
  switch in Escherichia coli.}
\newblock {\em Nature} 403(6767):339--342.

\bibitem{Elowitz2000}
Elowitz MB, Leibler S (2000) {A synthetic oscillatory network of
  transcriptional regulators.}
\newblock {\em Nature} 403(6767):335--338.

\bibitem{Basu2005}
Basu S, Gerchman Y, Collins CH, Arnold FH, Weiss R (2005) {A synthetic
  multicellular system for programmed pattern formation.}
\newblock {\em Nature} 434(7037):1130--4.

\bibitem{Sohka2009}
Sohka T, et~al. (2009) {An externally tunable bacterial band-pass filter.}
\newblock {\em Proc. Natl. Acad. Sci. U. S. A.} 106(25):10135--40.

\bibitem{Ajo-franklin2007}
Ajo-Franklin CM, et~al. (2007) {Rational design of memory in eukaryotic cells
  service Rational design of memory in eukaryotic cells}.
\newblock {\em Genes Dev.} 21(617):2271--2276.

\bibitem{Siuti2013}
Siuti P, Yazbek J, Lu TK (2013) {Synthetic circuits integrating logic and
  memory in living cells.}
\newblock {\em Nat. Biotechnol.} 31(5):448--52.

\bibitem{Daniel2013}
Daniel R, Rubens JR, Sarpeshkar R, Lu TK (2013) {Synthetic analog computation
  in living cells}.
\newblock {\em Nature} 497(7451):619--623.

\bibitem{Rubens2016}
Rubens JR, Selvaggio G, Lu TK (2016) {Synthetic mixed-signal computation in
  living cells}.
\newblock {\em Nat. Commun.} 7(7430):11658.

\bibitem{Xie2016}
Xie M, et~al. (2016) $\beta$-cell-mimetic designer cells provide closed-loop
  glycemic control.
\newblock {\em Science} 354(6317):1296--1301.

\bibitem{Jia2017}
Jia D, et~al. (2017) {Operating principles of tristable circuits regulating
  cellular differentiation}.
\newblock {\em Phys. Biol.} 14(3):035007.

\bibitem{DelVecchio:2008gy}
Del~Vecchio D, Ninfa AJ, Sontag ED (2008) {Modular cell biology: retroactivity
  and insulation.}
\newblock {\em Molecular Systems Biology} 4:161.

\bibitem{Jayanthi:2013do}
Jayanthi S, Nilgiriwala KS, Del~Vecchio D (2013) {Retroactivity Controls the
  Temporal Dynamics of Gene Transcription}.
\newblock {\em ACS synthetic biology} 2(8):431--441.

\bibitem{Tan:2009de}
Tan C, Marguet P, You L (2009) {Emergent bistability by a growth-modulating
  positive feedback circuit.}
\newblock {\em Nature chemical biology} 5(11):842--848.

\bibitem{Prindle:2014hj}
Prindle A, et~al. (2014) {Rapid and tunable post-translational coupling of
  genetic circuits.}
\newblock {\em Nature} 508(7496):387--391.

\bibitem{Ingram2006}
Ingram PJ, Stumpf MPH, Stark J (2006) {Network motifs: structure does not
  determine function.}
\newblock {\em BMC Genomics} 7:108.

\bibitem{Jimenez2017}
Jim{\'{e}}nez A, Cotterell J, Munteanu A, Sharpe J (2017) {A spectrum of
  modularity in multi-functional gene circuits}.
\newblock {\em Mol. Syst. Biol.} 13:925.

\bibitem{Purcell2011}
Purcell O, di~Bernardo M, Grierson CS, Savery NJ (2011) {A multi-functional
  synthetic gene network: A frequency multiplier, oscillator and switch}.
\newblock {\em PLoS One} 6(2).

\bibitem{Cotterell2010}
Cotterell J, Sharpe J (2010) {An atlas of gene regulatory networks reveals
  multiple three-gene mechanisms for interpreting morphogen gradients}.
\newblock {\em Mol. Syst. Biol.} 6(425):1--14.

\bibitem{Woods2016}
Woods ML, Leon M, Perez-Carrasco R, Barnes CP (2016) {A Statistical Approach
  Reveals Designs for the Most Robust Stochastic Gene Oscillators}.
\newblock {\em ACS Synth. Biol.} 5(6):459--470.

\bibitem{Otero-Muras2016}
Otero-Muras I, Banga JR (2016) {Design Principles of Biological Oscillators
  through Optimization: Forward and Reverse Analysis}.
\newblock {\em PLoS One} 11(12):1--26.

\bibitem{Strelkowa2010}
Strelkowa N, Barahona M (2010) {Switchable genetic oscillator operating in
  quasi-stable mode}.
\newblock {\em J. R. Soc. Interface} 7(48):1071--1082.

\bibitem{Suel2006}
S{\"{u}}el GM, Garcia-Ojalvo J, Liberman LM, Elowitz MB (2006) {An excitable
  gene regulatory circuit induces transient cellular differentiation.}
\newblock {\em Nature} 440(7083):545--50.

\bibitem{Jaeger2014}
Jaeger J, Monk N (2014) {Bioattractors: dynamical systems theory and the
  evolution of regulatory processes.}
\newblock {\em J. Physiol.} 592(Pt 11):2267--81.

\bibitem{Verd2014}
Verd B, Crombach A, Jaeger J (2014) {Classification of transient behaviours in
  a time-dependent toggle switch model.}
\newblock {\em BMC Syst. Biol.} 8(1):43.

\bibitem{Schaerli2014}
Schaerli Y, et~al. (2014) {A unified design space of synthetic stripe-forming
  networks.}
\newblock {\em Nat. Commun.} 5(May):4905.

\bibitem{Purcell2010}
Purcell O, Savery NJ, Grierson CS, di~Bernardo M (2010) {A comparative analysis
  of synthetic genetic oscillators.}
\newblock {\em J. R. Soc. Interface} 7(52):1503--1524.

\bibitem{Chau2012}
Chau AH, Walter JM, Gerardin J, Tang C, Lim WA (2012) {Designing synthetic
  regulatory networks capable of self-organizing cell polarization}.
\newblock {\em Cell} 151(2):320--332.

\bibitem{Carrera:2011fd}
Carrera J, Rodrigo G, Singh V, Kirov B, Jaramillo A (2011) {Empirical model and
  in vivo characterization of the bacterial response to synthetic gene
  expression show that ribosome allocation limits growth rate}.
\newblock {\em Biotechnology journal} 6(7):773--783.

\bibitem{Mather:2013hn}
Mather WH, Hasty J, Tsimring LS, Williams RJ (2013) {Translational cross talk
  in gene networks.}
\newblock {\em Biophysical Journal} 104(11):2564--2572.

\bibitem{Cookson:2011il}
Cookson NA, et~al. (2011) {Queueing up for enzymatic processing: correlated
  signaling through coupled degradation.}
\newblock {\em Molecular Systems Biology} 7:561.

\bibitem{Klumpp:2009vu}
Klumpp S, Zhang Z, Hwa T (2009) {Growth Rate-Dependent Global Effects on Gene
  Expression in Bacteria}.
\newblock {\em Cell}.

\bibitem{Scott:2010cx}
Scott M, Gunderson CW, Mateescu EM, Zhang Z, Hwa T (2010) {Interdependence of
  cell growth and gene expression: origins and consequences.}
\newblock {\em Science} 330(6007):1099--1102.

\bibitem{Cardinale:2013bt}
Cardinale S, Joachimiak MP, Arkin AP (2013) {Effects of genetic variation on
  the E. coli host-circuit interface.}
\newblock {\em Cell reports} 4(2):231--237.

\bibitem{Ceroni:2015ep}
Ceroni F, Algar R, Stan GB, Ellis T (2015) {Quantifying cellular capacity
  identifies gene expression designs with reduced burden.}
\newblock {\em Nature Methods} 12(5):415--418.

\bibitem{Borkowski:2016jz}
Borkowski O, Ceroni F, Stan GB, Ellis T (2016) {Overloaded and stressed:
  whole-cell considerations for bacterial synthetic biology.}
\newblock {\em Current opinion in microbiology} 33:123--130.

\bibitem{Verd2016}
Verd B (2016) {\em {EvoDevo in Phase Space: the dynamics of gap gene
  expression}} Ph.D. thesis.

\bibitem{Mathur2017}
Mathur M, Xiang JS, Smolke CD (2017) {Mammalian synthetic biology for studying
  the cell}.
\newblock {\em J. Cell Biol.} 216(1):73--82.

\bibitem{Panovska-Griffiths2013}
Panovska-Griffiths J, Page KM, Briscoe J (2013) {A gene regulatory motif that
  generates oscillatory or multiway switch outputs.}
\newblock {\em J. R. Soc. Interface} 10(79):20120826.

\bibitem{Balaskas2012}
Balaskas N, et~al. (2012) {Gene regulatory logic for reading the Sonic Hedgehog
  signaling gradient in the vertebrate neural tube.}
\newblock {\em Cell} 148(1-2):273--84.

\bibitem{Sokolowski2012}
Sokolowski TR, Erdmann T, ten Wolde PR (2012) {Mutual Repression Enhances the
  Steepness and Precision of Gene Expression Boundaries}.
\newblock {\em PLoS Comput. Biol.} 8(8).

\bibitem{Song2010}
Song C, et~al. (2010) {Estimating the stochastic bifurcation structure of
  cellular networks}.
\newblock {\em PLoS Comput. Biol.} 6(3).

\bibitem{Tian2006}
Tian T, Burrage K (2006) {Stochastic models for regulatory networks of the
  genetic toggle switch.}
\newblock {\em Proc. Natl. Acad. Sci. U. S. A.} 103(22):8372--8377.

\bibitem{Perez-Carrasco2016}
Perez-Carrasco R, Guerrero P, Briscoe J, Page KM (2016) {Intrinsic Noise
  Profoundly Alters the Dynamics and Steady State of Morphogen-Controlled
  Bistable Genetic Switches}.
\newblock {\em PLoS Comput. Biol.} 12(10):1--23.

\bibitem{Frigola2012}
Frigola D, Casanellas L, Sancho JM, Iba{\~{n}}es M (2012) {Asymmetric
  stochastic switching driven by intrinsic molecular noise}.
\newblock {\em PLoS One} 7(2).

\bibitem{Buzzi2015}
Buzzi CA, Llibre J (2015) {Hopf bifurcation in the full repressilator
  equations}.
\newblock {\em Math. Methods Appl. Sci.} 38(7):1428--1436.

\bibitem{BUSE2009}
Buşe O, Kuznetsov A, P{\'{e}}rez RA (2009) {Existence of limit cycles in the
  repressilator euqations}.
\newblock {\em Int. J. Bifurc. Chaos} 19(12):4097--4106.

\bibitem{Liepe2010}
Liepe J, et~al. (2010) {ABC-SysBio--approximate Bayesian computation in Python
  with GPU support.}
\newblock {\em Bioinformatics} 26(14):1797--9.

\bibitem{Liepe2014}
Liepe J, et~al. (2014) {A framework for parameter estimation and model
  selection from experimental data in systems biology using approximate
  Bayesian computation.}
\newblock {\em Nat. Protoc.} 9(2):439--56.

\bibitem{Gillespie2000}
Gillespie DT (2000) {The chemical Langevin equation}.
\newblock {\em J. Chem. Phys.} 113(1):297.

\bibitem{Clewley2012}
Clewley R (2012) {Hybrid Models and Biological Model Reduction with PyDSTool}.
\newblock {\em PLoS Comput. Biol.} 8(8):e1002628.

\bibitem{DeLaCruz2017}
de~la Cruz R, Perez-Carrasco R, Guerrero P, Alarcon T, Page KM (2017) {Minimum
  Action Path theory reveals the details of stochastic biochemical transitions
  out of oscillatory cellular states}.
\newblock {\em arXiv Prepr. arXiv1705.08683}.

\bibitem{Levine2013}
Levine JH, Lin Y, Elowitz MB (2013) {Functional roles of pulsing in genetic
  circuits.}
\newblock {\em Science} 342(6163):1193--200.

\bibitem{Lindner2004}
Lindner B, Garc{\'{i}}a-Ojalvo J, Neiman A, L SG (2004) {Effects of noise in
  excitable systems}.
\newblock {\em Phys. Rep.} 392(6):321--424.

\bibitem{Formosa-Jordan2012}
Formosa-Jordan P, Ibanes M, Ares S, Frade JM (2012) {Regulation of neuronal
  differentiation at the neurogenic wavefront}.
\newblock {\em Development} 139(13):2321--2329.

\bibitem{Kobayashi:2004cv}
Kobayashi H, et~al. (2004) {Programmable cells: interfacing natural and
  engineered gene networks.}
\newblock {\em Proceedings of the National Academy of Sciences of the United
  States of America} 101(22):8414--8419.

\bibitem{Kuznetsov:2006gx}
Kuznetsov A, Kaern M, Kopell N (2006) {Synchrony in a Population of
  Hysteresis-Based Genetic Oscillators}.
\newblock {\em SIAM Journal on Applied Mathematics} 65(2):392--425.

\bibitem{Danino2010}
Danino T, Mondrag{\'{o}}n-Palomino O, Tsimring L, Hasty J (2010) {A
  synchronized quorum of genetic clocks}.
\newblock {\em Nature} 463(7279):326--330.

\bibitem{Nikolaev:2016cv}
Nikolaev EV, Sontag ED (2016) {Quorum-Sensing Synchronization of Synthetic
  Toggle Switches: A Design Based on Monotone Dynamical Systems Theory.}
\newblock {\em PLoS Computational Biology} 12(4):e1004881.

\bibitem{Potvin-Trottier2016}
Potvin-Trottier L, Lord ND, Vinnicombe G, Paulsson J (2016) {Synchronous
  long-term oscillations in a synthetic gene circuit}.
\newblock {\em Nature} 538(7626):514--517.

\bibitem{Liu2017}
Liu J, et~al. (2017) {Coupling between distant biofilms and emergence of
  nutrient time-sharing.}
\newblock {\em Science} 356(6338):638--642.

\bibitem{Hilborn2012}
Hilborn RC, Brookshire B, Mattingly J, Purushotham A, Sharma A (2012) {The
  transition between stochastic and deterministic behavior in an excitable gene
  circuit}.
\newblock {\em PLoS One} 7(4).

\bibitem{Niederholtmeyer2015}
Niederholtmeyer H, et~al. (2015) {Rapid cell-free forward engineering of novel
  genetic ring oscillators.}
\newblock {\em Elife} 4(OCTOBER2015):e09771.

\bibitem{Izhikevich2000}
Izhikevich EM (2000) {Neural excitability, spiking and bursting}.
\newblock {\em Int. J. Bifurc. Chaos} 10(6):1171--1266.

\bibitem{Hubaud2014}
Hubaud A, Pourqui{\'{e}} O (2014) {Signalling dynamics in vertebrate
  segmentation}.
\newblock {\em Nat. Rev. Mol. Cell Biol.} 15(11):709--721.

\bibitem{Webb2016}
Webb AB, et~al. (2016) {Persistence, period and precision of autonomous
  cellular oscillators from the zebrafish segmentation clock}.
\newblock {\em Elife} 5(FEBRUARY2016):1--17.

\bibitem{Clark2017}
Clark E (2017) {Dynamic patterning by the Drosophila pair-rule network
  reconciles long-germ and short-germ segmentation}.
\newblock {\em bioRxiv} January:1--94.

\end{thebibliography}
 
\clearpage
\newpage      

\section*{Supporting Information}

\subsection*{Non-dimensional equations}

The nondimensional equations [\ref{eq:hillfunction}] result from the dimensional Hill function regulation,

\begin{eqnarray}
\frac {\mathrm{d}\tilde X}{\mathrm{d}\tilde t}&=&\frac{\tilde \alpha_X+\tilde \beta_X S}{1+S+(\tilde Z/\tilde z_X)^{n_{zx}}}- \tilde\delta_X \tilde{X} \nonumber\\
\frac {\mathrm{d}\tilde Y}{\mathrm{d}\tilde t}&=&\frac{\tilde \alpha_Y+\tilde \beta_Y S}{1+S+(\tilde X/\tilde x_Y)^{n_{xy}}}-\tilde\delta_Y \tilde{Y} \label{eq:hillfunctiondim}\\
\frac {\mathrm{d}\tilde Z}{\mathrm{d}\tilde t}&=&\frac{\tilde \alpha_Z}{1+(\tilde{X}/\tilde x_Z)^{n_{xz}}+(\tilde{Y}/\tilde y_Z)^{n_{yz}}}-\tilde\delta_Z Z. \nonumber
\end{eqnarray}

Measuring time in units of the degradation rate of protein $X$, all the temporal variables can be nondimensionalized as,

\begin{equation}
\delta_Y= \frac{\tilde\delta_Y}{\tilde \delta_X},\quad  \delta_Z= \frac{\tilde\delta_Z}{\tilde \delta_X},\quad t = \tilde t \delta_X.
\end{equation}

Similarly, concentrations and rates can be non-dimensionalized using the timescale of $\tilde \delta_X$ and the production rate $\tilde \alpha_Z$.

\begin{eqnarray}
&\alpha_X = \frac{\tilde \alpha_X}{\tilde \alpha_Z},\, \alpha_Y = \frac{\tilde \alpha_Y}{\tilde \alpha_Z},\, \beta_X = \frac{\tilde \beta_X}{\tilde \alpha_Z},\, \beta_Y = \frac{\tilde \beta_Y}{\tilde \alpha_Z}& \\
&z_x = \frac{\tilde z_x\tilde \delta_X}{\tilde \alpha_Z},\,x_y = \frac{\tilde x_y\tilde \delta_X}{\tilde \alpha_Z},\,x_z = \frac{\tilde x_z\tilde \delta_X}{\tilde \alpha_Z},\,y_z = \frac{\tilde y_z\tilde \delta_X}{\tilde \alpha_Z}& \\
&X = \frac{\tilde X\tilde \delta_X}{\tilde \alpha_Z},\,Y = \frac{\tilde Y\tilde \delta_X}{\tilde \alpha_Z},\,Z = \frac{\tilde Z\tilde \delta_X}{\tilde \alpha_Z}&
\end{eqnarray}

The signal $S$ is also measured in arbitrary units. Since $S$ is a control parameter to control the dynamics properties of the system, the results will hold for any non-linear relationship between concentration of inducer and $S$. 

For the stochastic Chemical Langevin Equation, the parameter $\Omega$ relates the non-dimensional expression levels with actual number of proteins ($N_X$, $N_Y$, $N_Z$) as,
\begin{equation}
N_X = \frac{X\tilde \alpha_Z\Omega}{\tilde \delta_X},\quad N_Y = \frac{Y\tilde \alpha_Z\Omega}{\tilde \delta_X},\quad N_Z = \frac{Z\tilde \alpha_Z\Omega}{\tilde \delta_X}.
\end{equation}

\subsection*{Parameter fitting}

The parameter exploration was carried out using Bayesian sampling techniques through the Approximate Bayesian Computation (ABC) using ABC-SysBio software \cite{Liepe2010}. The score functions, $d()$, designed are minimal for the optimal behavior scored. They were designed to capture a change from stable steady state to oscillations. This was evaluated on trajectories for each parameter set under the induction of two different signal values $S_{DC}$ and $S_{AC}$. 

First the network is induced by a low signal ($S=S_{DC}$) for $\Delta t = 50$. Allowing a transient of $\Delta t = 30$ (see Fig \ref{fig:scorefunction}), after which a constant response in time is scored:

\begin{equation}
d_{DC}(X(t)) = M_{DC} + 2 \frac{\max(X(t)) - \min(X(t))}{\max(X(t)) + \min(X(t))}
\end{equation}

Where $M_{DC}$ is the number of minima found, penalising oscillations. The second term of $d_{DC}$ penalises transients far from a steady expression.

The constant regime is perturbed by increasing the signal to a new value $S_{AC}=S_{DC}\sigma$ ($\sigma>1$) applied for $\Delta t = 100$. During this second period, the goodness of the oscillations was evaluated favouring large oscillation amplitudes, and penalising a non-constant amplitude in time:

\begin{equation}
d_{AC}(X(t)) = \left\lbrace\begin{array}{cc}
\frac{1}{M_{AC}}+2& M_{AC}<4\\
\left| \frac{A_M-A_{M-1}}{A_{M-1}}\right|+ 2 \frac{\min(X(t))}{\max(X{t}+\min(X))} & M_{AC}<4,
\end{array} \right.
\end{equation}

where $M_{AC}$ is the number of maxima found after a transient of $\Delta t =20$ region, and $A_M$ and $A_{M-1}$ are the amplitudes of the last and the previous to last full oscillations (see Fig. \ref{fig:scorefunction}).  Both parameters $S_{AC}$ and $\sigma$ were treated as free parameters of the optimisation.

Finally, in order to reduce artefacts arising from the choice of high Hill exponents, all the Hill exponents were fixed to $n=2$, varying only one exponent that is penalised to have higher values in circuits that already have a low score,

\begin{equation}
d_{Hill}(n_i) = \left\lbrace\begin{array}{cc}
2& d_{AC}+d_{DC}>2\\
\left| \frac{2-n_i}{3}\right|& d_{AC}+d_{DC}\leq2
\end{array} \right.
\end{equation}

The distance used to infer the parameters used in the the current study (Table \ref{tab:parameters}, and Fig. \ref{fig:histograms}) was,
 \begin{equation}
d = d_{DC}(X) + d_{AC}(X) + d_{Hill}(n_{zx}),
 \end{equation}
where the minimisation was run for 20 generations of the ABC optimisation and the expected behavior started to arise beyond generation 10 of the ABC optimisation. To test possible overfitting resulting from the functions used, alternative functions were designed resulting in similar results analysed during different generations of the algorithm, some examples are shown in Table \ref{tab:distances} where,
\begin{eqnarray}
d_1 &=& d^2_{DC}(Y)+d^2_{AC}(Y)+d^2_{Hill}(n_{zx}),\\
d_2 &=& d_{DC}(X) + d_{AC}(Y)+d_{Hill}(n_{yz}).
\end{eqnarray}

\subsection*{Stochastic expression}

The stochastic dynamics of expression is studied through the Chemical Langevin equation resulting from taking into account the stochastic nature of the production and degradation events \cite{Gillespie2000} as:

\begin{eqnarray}
\dot X&=&f_X(Z,S)- X + \sqrt{f_X^2(Z,S) + X^2}\xi_X(t) , \label{eq:CLE}\\
\dot Y&=&f_Y(X,S)-\delta_Y Y+ \sqrt{f_Y^2(X,S) + \delta_Y^2 Y^2}\xi_Y(t), \nonumber\\
\dot Z&=&f_Z(X,Y)-\delta_Z Z+ \sqrt{f_Z^2(X,Y) + \delta_Z^2 Z^2}\xi_Z(t), \nonumber
\end{eqnarray}

where $f_X$, $f_Y$, and $f_Z$ are the production terms of eqs.[\ref{eq:hillfunction}] and $\xi_i$ are uncorrelated white Gaussian noises of zero mean and autocorrelation $\langle \xi_i(t)\xi_i(t')\rangle = \Omega^{-1}\delta(t-t')$, where $\delta(t-t')$ is Dirac's delta and $\Omega$ is the system volume, relating expression concentrations with number of proteins. 

\subsection*{Cell lattice diffusion}

Spatially extended simulations for the genetic expression propagation were carried out implementing an array of hexagonal cell of unit length. One or more of the proteins forming the AC-DC circuit are allowed to diffuse between neighbours using a discrete Laplacian that for gene $X$  of the $i$-th cell reads

\begin{equation}
\dot X_i = D (\langle X \rangle_{\{ i \}}-X_i) ,
\end{equation}
where $D$ is the intercellular diffusion coefficient and  $\langle \cdot \rangle_{ \{ i \} }$ stands for the average expression of the target gene among of all the neighbouring cells of cell $i$.

%\showmatmethods % Display the Materials and Methods section

\newpage

\renewcommand\thefigure{S.\arabic{figure}}
\setcounter{figure}{0}    
\renewcommand\thetable{S.\arabic{table}}
\setcounter{table}{0}  

\subsection*{SI Movies}

\indent\begin{movie}
Expression of gene $X$ in a cellular lattice in time under a gradient of signal $S=10^{-7x/80-2}$. The gradient is equivalent to the signal set in Fig. \ref{fig:stability}, using the same color code. Initial condition is $X = 0, Y = 0, Z = 0$.
 \label{movie:phasecolor}
\end{movie}

\begin{movie}
Resulting spatiotemporal profile of Movie \ref{movie:phasecolor} separating the individual expression of each of the 3 genes X,Y and Z.
 \label{movie:phasechannels}
\end{movie}

\begin{movie}
Gene expression for a spatiotemporal signalling gradient in a cellular lattice. The signal starts at $S=0$ along the whole tissue and increases temporally building the same gradient as in Movie \ref{movie:phasecolor}. $S=10^{-7x/80-2}\mathrm e^{-2/t}$
 \label{movie:phasetimegradient}
\end{movie}

\begin{movie}
Time evolution of 6 different stochastic trajectories for the signalling $\Delta S_1$ for the same parameters and time protocol of Fig. \ref{fig:coherence}. The signal is fixed at S = 0.1 and is maintained until t=25 where it is increased gradually during $\Delta t = 2$ until reaching a final signal $S=100$ that is maintained until $t=100$. During that time the system undergoes a Hopf bifurcation setting the expression state close to the spiral centre. This entails a divergence between the different trajectories that results in a lack of coherence in the oscillations.
\label{movie:coherenceHopf} 
\end{movie}

\begin{movie}
Time evolution of 6 different stochastic trajectories for the signalling $\Delta S_2$ for the same parameters and time protocol as Fig. \ref{fig:coherence}. The signal is fixed at $S = 10^5$ and is maintained until t=25 where it is increased gradually during $\Delta t = 2$ until reaching a final signal $S=100$ that is maintained until $t=100$. The system undergoes a saddle-node bifurcation setting the expression state of the different trajectories at a similar position far from the limit cycle. The cycle attracts all the trajectories following the same slow manifold resulting in coherent convergence to the oscillatory behavior. 
\label{movie:coherenceSN} 
\end{movie}

\begin{movie}
Expression of gene Z in an array of cells where the AC-DC proteins can diffuse with a signal $S = 1000$ (bistable regime). Cells are initially at the constant expression steady state except one cell that is perturbed away from the steady state. That cell undergoes a cycle around the limit cycle increasing the levels of Z that diffuses to the neighbouring cells exciting them. The result is a propagating front of activation of gene Z.  $\Omega = 10^{5}$, $D = 0.1$ (see more details in SI)
\label{movie:propagationnice}
\end{movie}

\begin{movie}
Expression of gene Z in an array of cells where the AC-DC proteins can diffuse with a signal $S = 1000$ (bistable regime) a. Cells are initially at the constant expression steady state except one cell that is perturbed away from the steady state. A larger diffusion coefficient results in a wider and faster front than in  \ref{movie:propagationnice}. $\Omega = 10^{7}$, $D = 0.2$ (see more details in SI)\label{movie:propagationquick}
\end{movie}

\begin{movie}
Expression of gene Z in an array of cells where the AC-DC proteins can diffuse with a signal $S = 1000$ (bistable regime). Cells are initially at the constant expression steady state except one cell that is perturbed away from the steady state.  A larger intrinsic noise than \ref{movie:propagationnice} and \ref{movie:propagationquick} results in a front that leaves behind an excited media where intrinsic noise results in a dynamical spatial pattern. $\Omega = 2000$, $D = 0.2$ (see more details in SI)
\label{movie:propagationdirty}
\end{movie}

\begin{movie}
Expression of gene Z in an array of cells where the protein Z can diffuse intercellularly with a signal $S = 1000$ (bistable regime). Cells are initially at the constant expression steady state except one cell that is perturbed away from the steady state. Allowing only a single protein to diffuse still exhibits the excitable wave reducing its width and speed even considering a higher intrinsic noise and diffusion coefficient. $\Omega = 100$, $D = 0.5$ (see more details in SI)
\label{movie:propagationsingle}
\end{movie}

\clearpage

\subsection*{SI Figures and Tables}

\begin{figure}[h]
\includegraphics[width=1.\columnwidth]{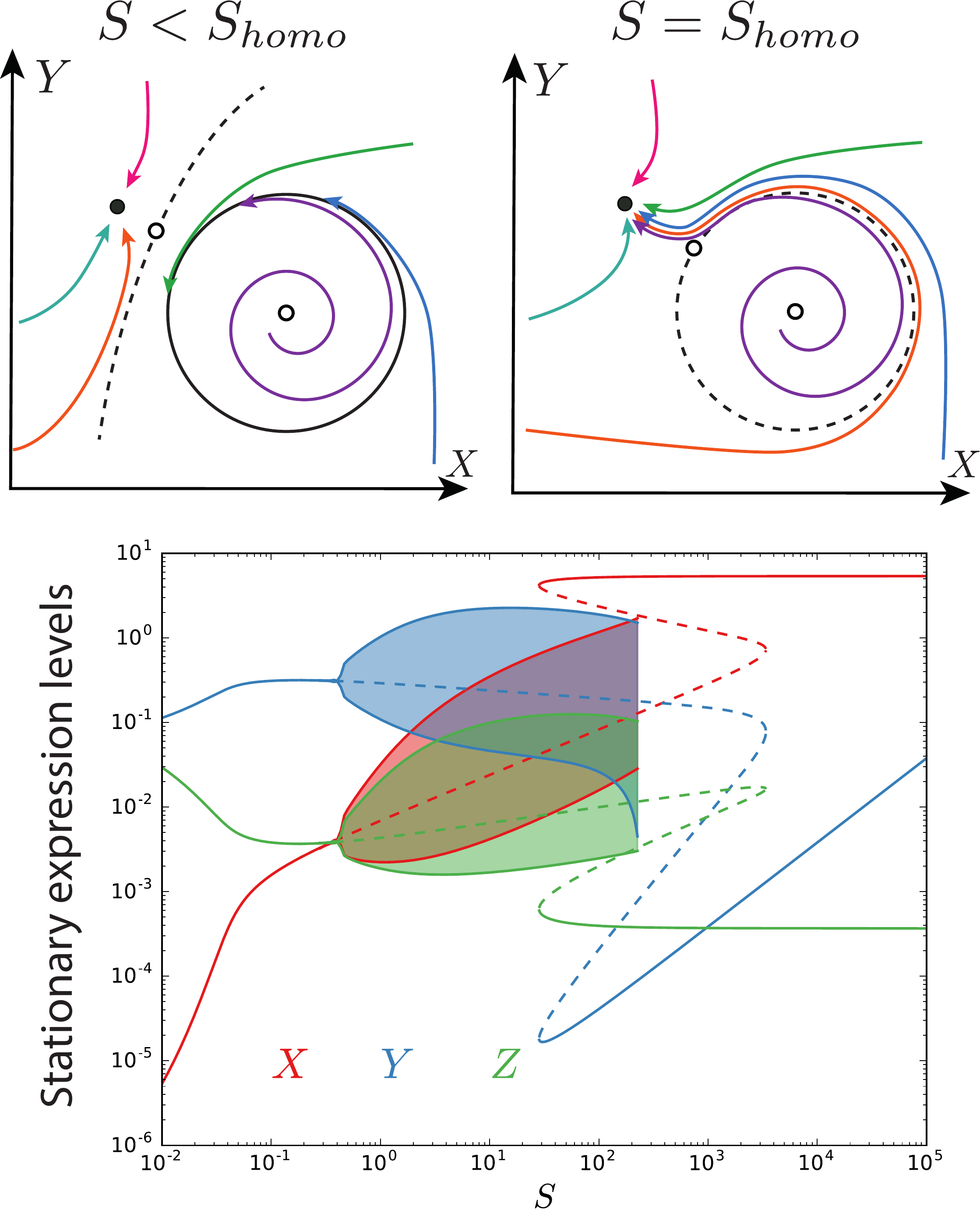}
\caption{\label{fig:homoclinic} Top) Scheme showing the attractor change at the homoclinic bifurcation. When the unstable saddle point (white point in the dashed line), collides with the limit cycle (solid circle), the cycle becomes unstable (dashed circle) and sustained oscillations are not available anymore. The only stable steady state remaining is the constant expression node that now attracts eventually all the expression trajectories except those starting precisely on the limit cycle. Bottom) Bifurcation diagram exhibiting a homoclinic bifurcation at $S\simeq200$.  Parameters used are those corresponding to $d^{(13)}$ in Table \ref{tab:distances}}
\end{figure} 

\begin{figure*}
\includegraphics[width=1\textwidth]{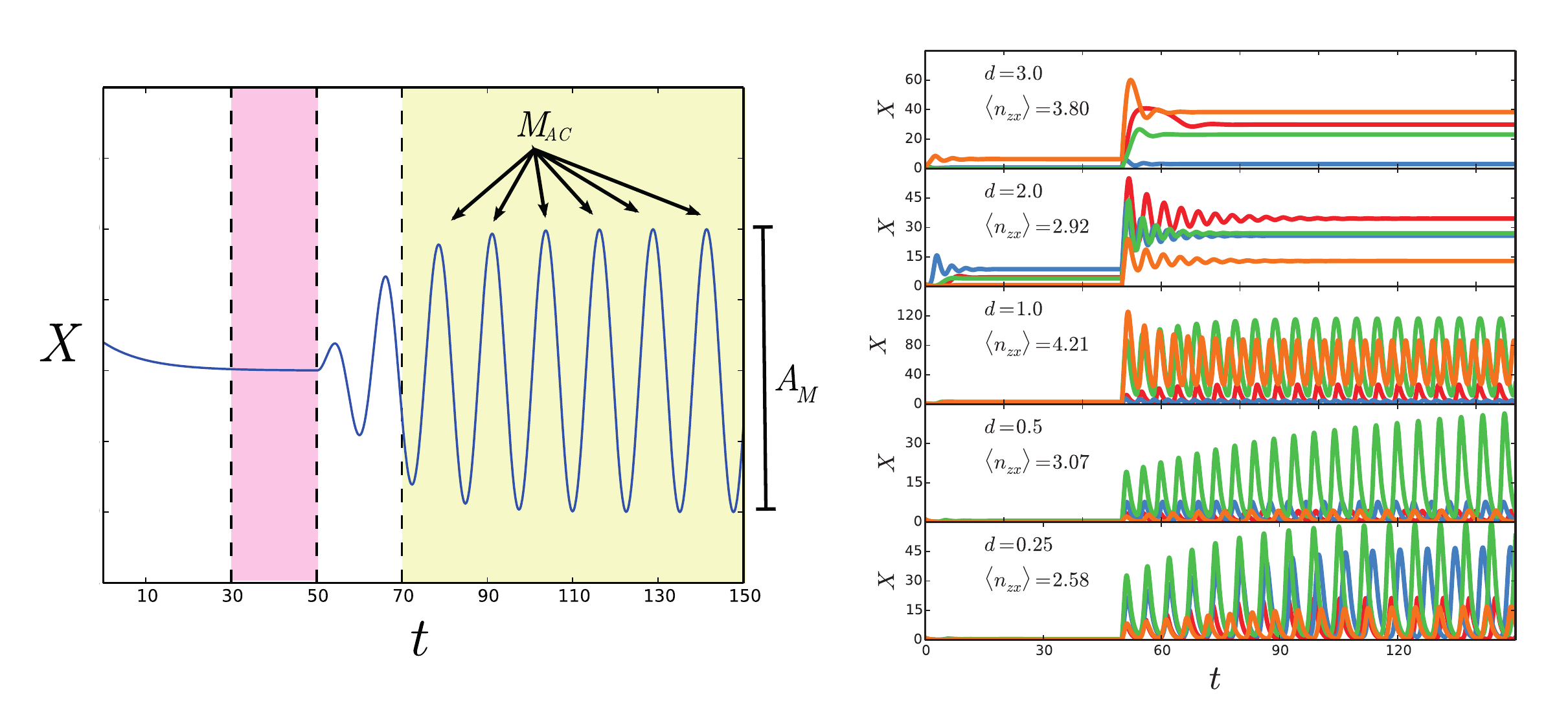}
\caption{\label{fig:scorefunction} Left) Target behavior used in the score function, evaluated at the shaded areas. The number of expression maxima at high signal $M_{AC}$ and amplitude of the last oscillations $A_M$ is used to compute the quality of oscillations. Right) Example results of behavior of the circuit under the two-signal protocol for different score results during the optimisation process. }
\end{figure*}

\begin{figure*}
\centering
\includegraphics[width=1.0\textwidth]{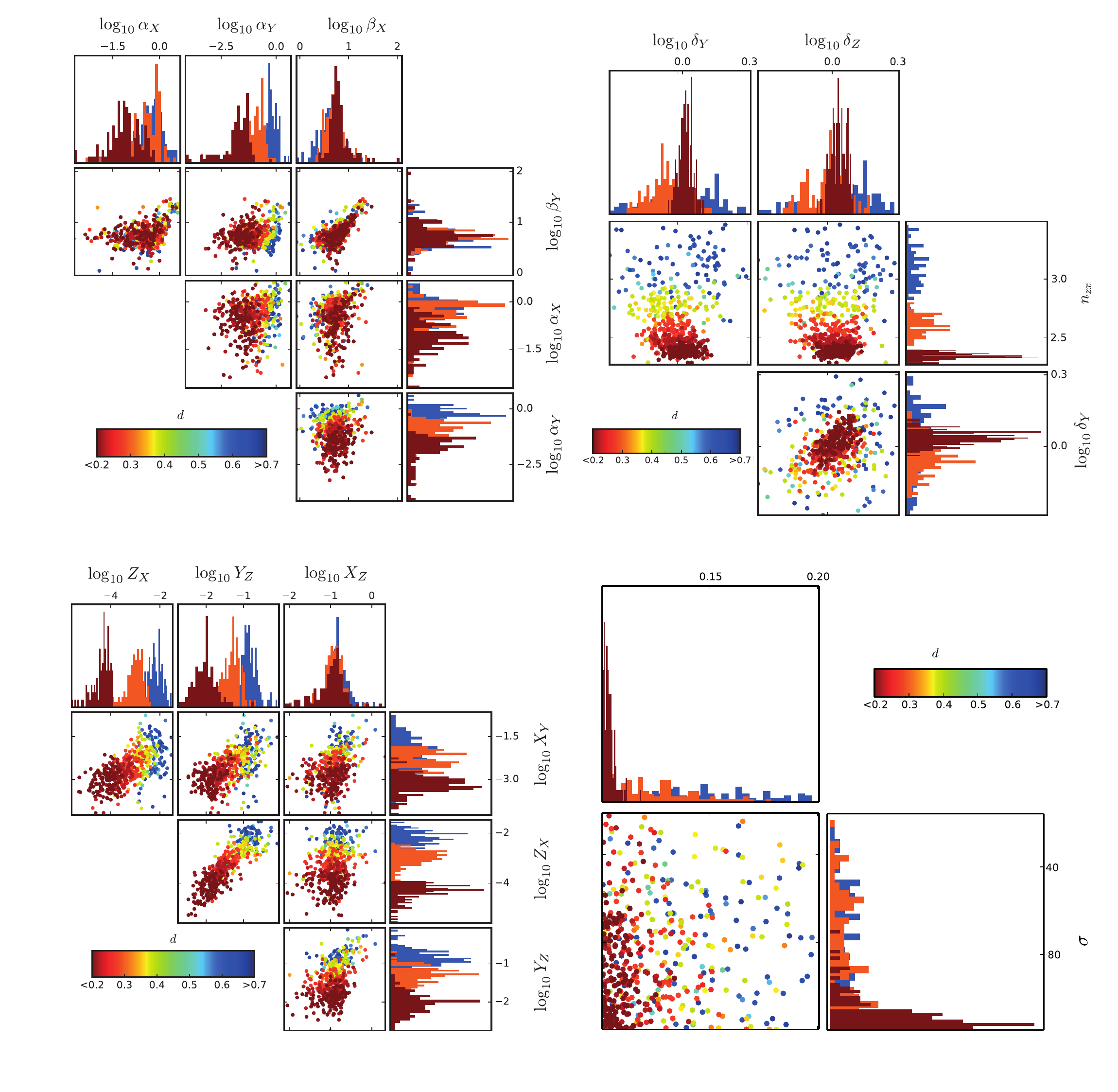}
\caption{\label{fig:histograms} Resulting parameter inference. Histograms correspond with the results of generations 10 (\emph{blue}), 15 (\emph{orange}) and 20 (\emph{dark red}), colored by the average distance score of the generation. Scatter plots contain the sample used in generations 5,8,11,14, and 17. Parameters are grouped in 4 categories to make easier the comparison between different magnitudes.}
\end{figure*}

\begin{figure*}
\centering
\includegraphics[width=0.8\textwidth]{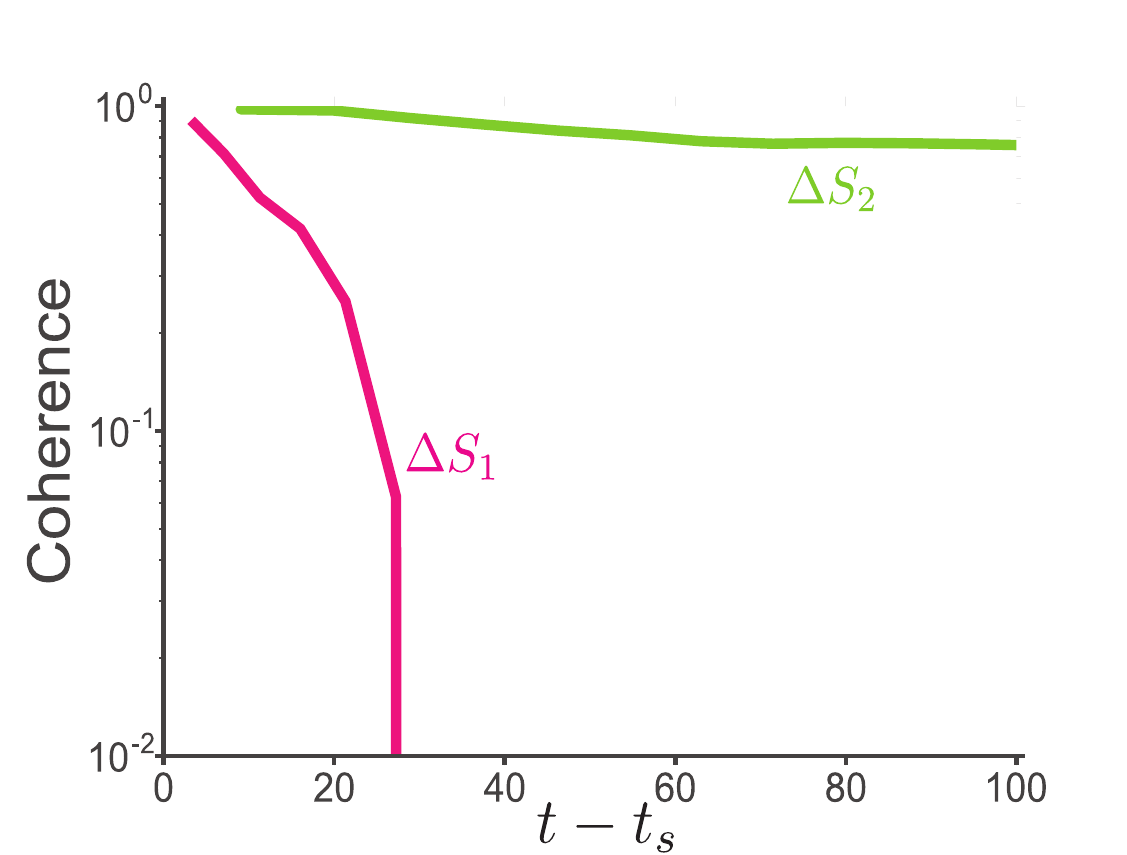}
\label{fig:coherencedetail}
\caption{ Comparison of the decrease in coherence in time for both signal histories $\Delta S_1$ and $\Delta S_2$ measured as $(\sigma_{max}-\sigma(t))/\sigma_{max}$, where $\sigma$ is the standard deviation of the phase of the oscillations for 20 simulations of each mechanism and $\sigma_{max}$ is the standard deviation corresponding to completely incoherent oscillations. Rest of parameters are the same as those of Fig. \ref{fig:coherence} .}
 \end{figure*}

\begin{table}[h]
%\rowcolors{2}{orange!25}{white}
\centering
\begin{tabular}{cccc}
Parameter&$\bar x\pm$ s.d&mode\\\hline\hline
$\alpha_X$&$(15\pm 14)10^{-2}$&$3.9\cdot 10 ^{-2}$\\
$\alpha_Y$&$(2.7\pm 2.1)10^{-2}$&$4.3 \cdot 10^{-3}$\\
$\beta_X$&$5.9\pm0.8$&$6.1$\\
$\beta_Y$&$5.4\pm1.2$&$5.7$\\
$\delta_Y$&$1.07\pm0.08$&$1.05$\\
$\delta_Z$&$1.12\pm0.09$&$1.04$\\
$z_X$&$(6.4\pm4.3)10^{-5}$&$1.3\cdot10^{-5}$\\
$y_Z$&$(11\pm 4)10^{-3}$&$11\cdot 10^{-3}$\\
$x_Z$&$(12\pm 5)10^{-2} $&$12\cdot10^{-2}$\\
$x_Y$&$(8.3\pm 4.2)10^{-4}$&$7.9\cdot10^{-4}$\\
$n_{zx}$&$2.34\pm0.04$&$2.32$\\
\end{tabular}
\caption{\label{tab:parameters} Optimal parameters resulting from the last 300 points of the ABC fitting (generations 18-20). Each row shows the mean, standard deviation, and position of the peak of the distribution (mode). For details see SI.}
\end{table}

\begin{table*}
\centering
%\rowcolors{2}{orange!25}{white}
\begin{tabular}{cccccc}
&$d$&$d^{(13)}$&$d^*$&$d_1$&$d_2$\\\hline\hline
$ \alpha_X$&$0.57\pm 0.37$&$0.43\pm0.31$&$0.64\pm 0.37$&$0.60\pm 0.39$&$0.60\pm 0.39$  \\
$ \alpha_Y$&$0.21\pm 0.19$&$0.066\pm0.055$&$0.18\pm 0.14$&$0.19\pm 0.14$&$0.23\pm 0.17$\\
$ \beta_X$&$4.9\pm 1.3$&$5.4\pm 1.1$&$5.1\pm1.5$&$4.9\pm 1.4$&$5.1\pm 1.3$\\
$ \beta_Y$&$5.7\pm 1.0$&$5.2\pm 1.3$&$5.7\pm 1.2$&$5.6\pm 1.1$&$5.5\pm 1.1$\\
$ \delta_Y$&$0.93\pm 0.18$ &$ 1.0 \pm 0.1$&$0.94\pm 0.17$&$1.02\pm 0.27$&$1.00\pm 0.27$\\
$ \delta_Z$&$1.04\pm 0.22$& $ 1.1 \pm 0.1$&$1.06\pm 0.21$&$1.05\pm 0.26$&$1.22\pm 0.27$\\
$z_X$&$(1.7\pm 1.4)10^{-3}$&$(2.7\pm2.2)10^{-4}$&$(1.6\pm 1.5)10^{-3}$&$(3.0\pm 2.1)10^{-3}$&$(6.0\pm 5.8)10^{-4}$\\
$y_Z$&$(6.4\pm 2.8)10{-2}$&$(2.0\pm0.9)10^{-2}$&$(6.1\pm 2.5)10^{-2}$&$(9.0\pm 3.8)10^{-2}$&$(4.6\pm 2.4)10^{-2}$\\
$x_Z$&$0.13\pm 0.05$&$0.11\pm0.05$&$0.14\pm 0.05$&$0.14\pm 0.04$&$0.13\pm 0.05$\\
$x_Y$&$(5.7\pm 4.5)10^{-3}$&$(1.8\pm1.3)10^{-3} $&$(4.8\pm 3.7)10^{-3}$&$(12.4\pm 8)10^{-3}$&$(11\pm 9)10^{-3}$\\
$n$&$2.70\pm 0.15$&$ 2.44\pm 0.07 $&$2.67\pm 0.14$&$2.42\pm 2.74$&$2.70\pm 0.16$\\
$\sigma$&$106\pm 44$&$ 135 \pm 39 $&$120\pm 42$&$114\pm 42$&$120\pm 40$\\
$ S_{DC}$&$0.13\pm 0.13$&$   0.22 \pm 0.01$&$0.024\pm 25$&$0.132\pm 0.023$&$0.136\pm 0.023$\\
\rowcolor{red!25}$d$&$0.33\pm 0.04$&$0.22\pm0.01$&$0.32\pm 0.03$&$0.088\pm 0.023$&$0.359\pm 0.047$
\end{tabular}
\caption{\label{tab:distances}\small  Inferred mean and standard deviation of the sampled parameter distribution for generations 8,9 and 10 (300 points). $d^*$ is a second realisation of the inference using $d$ to check the robustness of the optimisation. Parameter $n$ is the varying Hill exponent corresponding to each score distance. $d^{(13)}$ corresponds to generations 13, 14 and 15 for distance $d$. Alternative score functions$d_1$ and $d_2$ return similar parameter relationships.}
\end{table*}

% \pnasbreak splits and balances the columns before the references.
% If you see unexpected formatting errors, try commenting out this line
% as it can run into problems with floats and footnotes on the final page.
%\pnasbreak

\end{document}